\documentstyle[preprint,aps]{revtex}
\tightenlines
\begin{document}
\draft
\author{A.Gammal$^1$ and T.Frederico$^2$}
\address{$^1$FINPE, Instituto de F\'{\i}sica, Universidade\\
de S\~{a}o Paulo, CP66318, 05315-970, S\~{a}o Paulo, Brazil \\
$^2$ Departamento de F\'{\i}sica, Instituto Tecnol\'ogico da Aeron\'autica, 
\\
Centro T\'ecnico Aeroespacial, 12228-900 S\~ao Jos\'e dos Campos,\\
S\~ao Paulo, Brazil}
\title{The Nuclear Sigma Term in the Skyrme Model: Pion-Nucleus Interaction}
\maketitle

\begin{abstract}
The nuclear sigma term is calculated including the nuclear matrix element of
the derivative of the NN interaction with respect to the quark mass,
$m_q\frac{\partial V_{NN}}{\partial m_q}$. The NN potential is evaluated in the
skyrmion-skyrmion picture within the quantized product ansatz. The
contribution of the NN potential to the nuclear sigma term provides
repulsion to the pion-nucleus interaction. The strength of the s-wave
pion-nucleus optical potential is estimated including such contribution. The
results are consistent with the analysis of the experimental data.
\end{abstract}



\section{Introduction}

Nowadays Quantum Chromodynamics is accepted as the theory for strongly
interacting particles. It was probed with success at high energies where
perturbative expansion can be used. Unfortunately in the low energy regime
the perturbative expansion is not valid and nonperturbative schemes such as
lattice QCD have been implemented, but still with enormous computational
efforts. The energy region in which  quarks and gluons 
build the effective hadronic degrees of freedom is of interest for nuclear
physics applications. Effective theories have been constructed considering only
hadronic degrees of freedom and designed to work at low energies. In such
models the main guides are the symmetries of the fundamental theory. A
bridge can be made between the two descriptions when considering the QCD
vacuum structure. The presence of hadronic matter affects the vacuum
structure, which is reflected by the modification of the quark and gluon
condensates \cite{Cohen1}. The hadronic matter dig a hole in the local value
of the condensate as expressed by the expectation value of the local
composite operator $\overline q q $, where $q$ is the quark operator.

The difference between the volume integral of the quark condensate in the
presence of hadronic matter and its vacuum value is proportional to the
nuclear sigma term. It can be obtained by the application of the
Hellmann-Feynman theorem, which allows to express the nuclear sigma term as $%
m_q\frac{\partial M_A}{\partial m_q} $ with $M_A$ being the nuclear mass 
\cite{Cohen1}. We are limiting our discussion to the SU(2) flavor sector.
The nucleon-nucleon interaction is responsible for the nuclear binding, so
it is natural to  question  how the NN potential will be affected
by variations of the  quark masses, if a reliable calculation of the nuclear
sigma term is desired\cite{Cohen}.

However, to predict how the potential depends on the current quark masses of
the fundamental theory, the connection of the effective hadronic model with
QCD is necessary. The quark mass terms break explicitly the chiral symmetry
in the QCD Lagrangian. In the SU(2) flavor sector of the theory, the up-down
quark masses originate the pion mass, since the pion is considered the
Goldstone boson arising from the spontaneous breaking of SU(2)$_{\text{L}%
}\times $SU(2)$_{\text{R}}$ chiral symmetry \cite{gor}.

The Skyrme model allows to study the effect of the explicitly chiral
symmetry breaking in a chiral invariant hadronic model. The quark mass
comes through the pion mass term in the Skyrme Lagrangian. Such connection
is made using the Gell-Mann, Oakes and Renner (GOR) relation\cite{gor}, 
\begin{equation}
m_\pi ^2f_\pi ^2=-2m_q\left\langle \left| \bar qq\right| \right\rangle
_{vac}\ ,  \label{gor}
\end{equation}
where $m_q=(m_u+m_d)/2$, $m_u$ and $m_d$ are the up and down current quark
masses, respectively. The mass $m_q$ is quite small in the hadronic scale
and the local value of the vacuum condensate is $\left\langle \left| \bar q
q\right| \right\rangle _{vac}:= \left\langle \left| \bar uu + \bar dd\right|
\right\rangle _{vac}/2\ $.

The Skyrme Model \cite{Skyrme,Sky} describes qualitatively the nucleon \cite
{Adkins} and the nucleon-nucleon interaction \cite{Jackson,Vimau,Wamb}. The
particular feature of the model is that the pion mass term in the Lagrangian
contributes not only for the nucleon properties but also to the NN
potential. We are supposing that the dependences of the pion weak decay
constant ($f_\pi $) and the Skyrme parameter ($e$) on the up-down quark
masses are small and they are not considered here.

The derivative of any hadronic quantity, $O$, with the up-down average quark
mass in the Skyrme model is translated into a derivative in $m^2_\pi$ using
the GOR relation, Eq.(\ref{gor}), such that 
\begin{equation}
m_q\frac{\partial O}{\partial m_q}=m_\pi ^2\frac{\partial O}{\partial m_\pi^2%
}.  \label{oper}
\end{equation}

In this work, we obtain the derivative of the NN interaction with respect to
the current quark mass by calculating the potential with the product ansatz
approximation in the Skyrme model. We discuss how the derivative $m_q\frac{%
\partial V_{NN}}{\partial m_q}$ enters in the evaluation of the nuclear
sigma term in the context of nonrelativistic quantum mechanics and its
contribution to the quark condensate. The expectation value of $m_q\frac{%
\partial V_{NN}}{\partial m_q}$ in the nuclear state contributes to the
repulsive strength of the low-energy isoscalar pion-nucleus s-wave optical
potential. These calculations can be tested through the comparison with the
results of the pion-nucleus scattering and pionic atom data analysis \cite
{Sek83,Sek88,Sal95}. For completeness, we also calculate the nucleon sigma
term, the derivatives of the pion-nucleon coupling constant($g_{\pi NN})$
and axial charge ($g_A$).

In Section II, we discuss the nuclear sigma term in the context of
nonrelativistic quantum mechanics, defining the part of the sigma term
associated with the NN interaction. In Section III, we discuss the quark
condensate at finite nuclear density with the inclusion of the new term. In
Section IV, the Skyrme model is briefly discussed. In Section V, the results
for the derivative of the NN potential, with respect to the quark mass, are
presented together with the derivatives of the one-body observables. In
Section VI, part of the main results of this work are presented, those
correspond to the 
calculations of the nuclear sigma term for several nuclei. In
Section VII, the contribution of the derivative of the NN potential to the
pion s-wave optical potential is calculated and compared to the results of
the analysis of experimental data. In Section VIII, we summarize our
conclusions.

\section{ The Nuclear Sigma Term.}

The sigma term is a measure of the isoscalar content of the nuclear state ($%
\Psi _A$), with $A$ nucleons, and thus it gives the strength of chiral
symmetry breaking due to the $u$ and $d$ quark masses. It can be directly
defined as the expectation value of the commutators of the axial charge $%
(Q^5)$ with the Hamiltonian\cite{del,Cheng}, given by 
\begin{equation}
\Sigma _A=\left\langle \Psi _A\left| \left[ Q^5,\left[ {Q}^5,H\right]
\right] \right| \Psi _A\right\rangle .
\end{equation}
In QCD framework it can be expressed as \cite{Cheng} 
\begin{equation}
\Sigma _A=2m_q\int d^3x{(}\left\langle \Psi _A|\bar q(\vec x{)}q(\vec x{)}|\
\Psi _A\right\rangle -\left\langle 0|\bar q(\vec x{)}q(\vec x{)}%
|0\right\rangle )=m_q\frac{\partial M_A}{\partial m_q}\ ,  \label{snuc}
\end{equation}
it is proportional to the nonstrange quark condensate. The last equality in
Eq. (\ref{snuc}), follows from the Hellmann-Feynman theorem \cite{Cohen1} at
the level of the fundamental QCD theory, where the quark masses contribute
to the Hamiltonian as $\delta H_{QCD}=m_u\bar uu+m_d\bar dd$.

The nucleon sigma term is 
\begin{equation}
\Sigma _N=m_q\frac{\partial M_N}{\partial m_q}\ ,
\end{equation}
where we used (\ref{oper}) and $M_N$ is the nucleon mass. The experimental
value extracted from pion-nucleon scattering is  $\Sigma _N\sim 45$ MeV 
\cite{Gas91}.

The nuclear sigma term has a contribution from the derivative of the nuclear
binding energy ($E_B$) with respect to the quark mass. Using that $%
M_A=ZM_p+(A-Z)M_n+E_B$, where $Z$ is the proton number, $M_p$ and $M_n$ are
the proton and neutron masses, respectively, we have: 
\begin{equation}
\Sigma _A=A\Sigma _N+m_q\frac{\partial E_B}{\partial m_q}\ ,
\end{equation}
where the neutron and proton sigma terms were considered equal.

The effect of the nuclear binding in the sigma term can be calculated, for
example, from the Schr\"{o}dinger equation in the center of mass system. If
the quark masses contribute effectively to $V_{NN}$, they also influence the
binding energies in the Schr\"{o}dinger equation, written as 
\begin{equation}
H \left| \Psi_A\right\rangle = \left[ \sum_{i=1,A}-\frac{\hbar ^2}{2M_N}%
\nabla _i^2+ \frac{1}{2}\sum_{i \ne j} V_{ij}\right]\left|
\Psi_A\right\rangle = E_B \left|\Psi_A\right\rangle \ .  \label{schro}
\end{equation}

Using the Hellman-Feynman theorem and considering that $M_N$ and $V_{ij}$
depend on $m_q$, the nuclear sigma term is given by 
\begin{eqnarray}
\Sigma _A=A\Sigma _N+\Sigma _A^K+\Sigma _A^V\ ,  \label{sign}
\end{eqnarray}
where 
\begin{eqnarray}
\Sigma _A^V=\left\langle \Psi _A\right| \frac 12\sum_{i\neq j}m_q\frac{%
\partial V_{ij}}{\partial m_q}\left| \Psi _A\right\rangle \ ,  \label{sigv}
\end{eqnarray}
is the contribution of the NN potential to the nuclear sigma term, and 
\begin{eqnarray}
\Sigma _A^K=-\frac{\Sigma _N}{M_N}\overline{E}_K\ ,  \label{sigk}
\end{eqnarray}
is the term coming form the kinetic energy $\overline{E}_K$, which is the
expectation value of the sum of the nucleon kinetic energies in the center
of mass frame of the nucleus. The $A\Sigma _N$ term in Eq.(\ref{sign}) has
been discussed in the context of nuclear matter in Ref.\cite{Cohen1,Cohen}.
The last term in Eq. (\ref{sign}) can be rewritten using Eq.(\ref{oper}), as 
\begin{eqnarray}
\frac 12\sum_{i\neq j}m_q\frac{\partial V_{ij}}{\partial m_q}=\frac 12%
\sum_{i\neq j}m_\pi ^2\frac{\partial V_{ij}}{\partial m_\pi ^2}\ .
\label{dvpi}
\end{eqnarray}
The above equality permits to calculate the derivative $m_q\frac{\partial
V_{NN}}{\partial m_q}$ in the Skyrme model.

\section{The Quark Condensate}

The nuclear sigma term is responsible for the change in the quark condensate
at finite density \cite{Cohen1,Cohen}. The sigma term for nuclear matter at
a finite density ($\rho $) in a given volume ($Vol$) is given by 
\begin{equation}
\Sigma _\rho =2m_qVol{(}\left\langle \rho |\bar q(\vec x{)}q(\vec x{)}|\rho
\right\rangle -\left\langle 0|\bar q(\vec x{)}q(\vec x{)}|0\right\rangle
)=m_q\frac{\partial ({\cal E}Vol)}{\partial m_q}\text{ ,}  \label{3.47}
\end{equation}
where ${\cal E}$ is the nuclear matter energy density given by 
${\cal E}=M_N\rho +%
\frac{\overline{E}_V+\overline{E}_K}A\rho ,$ in the limit of $A\rightarrow
\infty $ and $\overline{E}_V$ is the potential energy.
Since $\rho =A/Vol,$ we get from (\ref{3.47}) 
\begin{equation}
2m_q\left( \left\langle \bar qq\right\rangle _\rho -\left\langle \bar q%
q\right\rangle _{vac}\right) =m_q\frac{\partial M_N}{\partial m_q}\rho +m_q%
\frac{\partial (\overline{E}_V/A+\overline{E}_K/A)}{\partial m_q}\rho \ .
\end{equation}
Using (\ref{sign}), the definitions $\Sigma _A^V$(\ref{sigv}), $\Sigma
_A^K $(\ref{sigk}) and the GOR relation, $2m_q\left\langle \bar q%
q\right\rangle _{vac}=-m_\pi ^2f_\pi ^2,$ we have 
\begin{equation}
\frac{\left\langle \bar qq\right\rangle _\rho }{\left\langle \bar q%
q\right\rangle _{vac}}\simeq 1-\frac{\Sigma _N+\left( \Sigma _A^V+\Sigma
_A^K\right) /A}{m_\pi ^2f_\pi ^2}\rho +... \ \ .  \label{3.49}
\end{equation}
If we neglect $\left( \Sigma _A^V+\Sigma _A^K\right) $ the expression above
reduces to the leading order term found in Ref. \cite{Cohen}. Recent works
by Li and Ko \cite{Ko94}, Delfino et al \cite{Del95} and Brockman and Weise 
\cite{BW96}, calculated the quark condensate in relativistic nuclear matter
theories following the work of Cohen and collaborators \cite{Cohen}. They
included in the evaluation of the nuclear sigma term the derivatives of the
meson masses with respect to the quark mass, and in their context the quark
condensate didn't deviate strongly from the leading order calculation
at normal density. Brown
and Rho \cite{BR96} have studied the quark condensate in nuclear medium
without the contribution of the NN potential to the nuclear sigma term. The
values of the condensate at normal density, in all cases, are around the
leading order result, $\left\langle \bar qq\right\rangle _{\rho
_N}/\left\langle \bar qq\right\rangle _{vac}\approx 0.70$ .

In the following section we will calculate $\Sigma _A^V+\Sigma _A^K$ and $%
\Sigma _A$ using the Skyrme Model and the GOR relation. Subsequently we will
calculate the contribution of $\Sigma _A^V$ to the pion-nucleus optical
potential. In the summary, we will present an estimate to the quark
condensate using (\ref{3.49}), with $\Sigma _A^V$ obtained from the Skyrme
model.

\section{ The Skyrme Model.}

The Skyrme Lagrangian\cite{Skyrme} can be expressed as 
\begin{equation}
{\cal L}=\frac{f_\pi ^2}4Tr\left( \partial _\mu U\partial ^\mu U^{\dagger
}\right) +\frac 1{32e^2}Tr\left[ U^{\dagger }\partial _\mu U,U^{\dagger
}\partial _\nu U\right] ^2+\frac 12f_\pi ^2m_\pi ^2(TrU-2)\ .  \label{lsu}
\end{equation}
where $U=e^{i{\vec \tau }.\vec \pi }$; ${\vec \tau }$ are the Pauli isospin
matrices. Assuming the hedgehog ansatz, ${\vec \pi }{\ (\vec r)}=\hat rF(r),$
and minimizing the energy we arrive at a differential equation for the
profile function $(F)$ 
\begin{equation}
\left( \frac{u^2}4+2s^2\right) F^{\prime \prime }+\frac 12uF^{\prime
}+F^{\prime 2}2sc-\frac 12sc-\frac{2s^3c}{u^2}-\beta ^2u^2\sin F=0\ ,
\end{equation}
where $u=2ef_\pi r$, $\beta =m_\pi /(2ef_\pi )$, $s=\sin F$ and $c=\cos F$ .
The solution with baryon number B=1, can be achieved imposing $F(0)=\pi $
and $F(\infty )=0.$

We make use of the quantized version \cite{Jackson,Vimau} of the product
ansatz to get a qualitative idea of the consequences of finite quark masses
in the NN potential. The product ansatz was originally introduced by Skyrme%
\cite{Sky} as an attempted solution for the skyrmion-skyrmion interaction.
It has the virtue of giving baryon number B=2 for any separation distance
and gives the one pion exchange potential for large distances. It provides
an NN potential which can be decomposed in the form 
\begin{equation}
V_{NN}=V_C+{\vec \tau }_{1.}{\vec \tau }_2\left[ V_{SS}\left( {\vec \sigma }%
_1.{\vec \sigma }_2\right) +V_TS_{12}\right] \ ,
\end{equation}
where $V_C$ is the isoscalar central potential, $V_{SS}$ is the spin-spin
potential and $V_T$ is the tensor potential; $S_{12}=3\left( {\vec \sigma }%
_1.{\ \hat r}\right) \left( {\vec \sigma }_2.{\hat r}\right) -{\vec \sigma }%
_1.{\vec \sigma }_2.$ ; ${\vec \sigma }$ are the Pauli spin matrices.

Exact numerical solutions for the skyrmion-skyrmion interaction were
obtained in the work of Walhout and Wambach\cite{Wamb}. In that work\cite
{Wamb}, they compared the exact numerical calculations to the product ansatz
result and to the Paris potential. Although the gross feature of the central
component is given by the product ansatz, it misses the attraction which is
present in the exact calculation and Paris potential. The product ansatz is
roughly reliable for $r\geq 1.5$ fm in the tensor and spin-spin channels.

Although the product ansatz in the NN central potential is only approximated
when compared to exact numerical calculations\cite{Wamb}, we believe that is
possible to use it in the calculation of the derivative. 
According to Jackson et al%
\cite{Jackson}, the difficulties in reproducing the NN interaction from the
product ansatz come from the fact that the potential is a difference of two
large quantities  of magnitudes two orders  higher than the potential itself,
and so little deformation in the skyrmions can originate drastic effects.
This shall not occur in the derivative of the potential with respect to the
quark mass, since the final result is of the same order of 
magnitude of the individual nucleon contributions. 
So, deformations that could occur in the skyrmions as
they are close, would be in this case only corrections that could not
change the magnitude of the result.

\section{ Derivatives of the Nucleon Observables and NN Potential}

We begin by showing the results of the derivatives of the one-body
observables with respect to the quark mass. We can calculate several nucleon 
observables \cite{Adkins}, such as the nucleon mass, the pion-nucleon 
coupling $(g_{\pi NN})$ and the axial coupling $(g_A)$, 
from the Skyrme Model in the semi-classical quantization approach.
Thus, by changing the pion mass we can
easily calculate $m_\pi ^2\frac{\partial M_N}{\partial m_\pi ^2}$, $m_\pi ^2%
\frac{\partial g_{\pi NN}}{\partial m_\pi ^2}$ and $m_\pi ^2\frac{\partial
g_A}{\partial m_\pi ^2}$ . These calculations provide the determination of
the sigma term, $m_q\frac{\partial g_{\pi NN}}{\partial m_q}$ and $m_q\frac{%
\partial g_A}{\partial m_q}$, respectively. The results of the derivatives
of the observables are presented in Table I, and for completeness, the value
of the observables themselves are also shown.

The calculations were performed with two sets of parameters, $f_\pi =54$MeV, 
$e=4.84$ \cite{Adkins} and $f_\pi =93$MeV, $e=4.0$ \cite{Holz} with $m_\pi
=138$MeV. In Table I, the values of the nucleon sigma term, were included.
It has been calculated previously in Ref.\cite{Adkins}. They obtained 38 MeV
and 49 MeV, for the set $f_\pi =$ 54 MeV and $e=$ 4.84, using different
methods without recurring to the Hellmann-Feynman theorem. These values are
comparable to our result of 52.2 MeV. The experimental value of about 45 MeV 
\cite{Gas91}, is somewhat consistent with our value of 59.6 MeV for $f_\pi =$
93 MeV and $e=$ 4.0 (see Table I). We will restrict our evaluation of the
nuclear sigma term to this set of parameters, since $f_\pi $ is taken from
the experimental value.

The coupling constant $g_{\pi NN}$ has very different 
derivatives for the two sets of
parameters, while the derivatives of $g_A$ are similar, as shown in Table I.
However, in the case of the $g_{\pi NN}$, the derivatives $m_q\frac{\partial
g_{\pi NN}}{\partial m_q}$, the nucleon sigma term and $m_q\frac{\partial g_A%
}{\partial m_q}$ are consistent with the Goldberger-Treiman relation \cite
{Cheng}.

The derivative $m_q\frac{\partial V_{NN}}{\partial m_q}$ is evaluated
numerically in the Skyrme model using the quantized product ansatz with the
parameter set $f_\pi =$ 93 MeV and $e=$ 4.0. We calculated the NN
interaction in skyrmion-skyrmion picture and recalculated it for a small
change in the pion mass. Thus, we obtained numerically the quantities $m_q%
\frac{\partial V_C}{\partial m_q}$, $m_q\frac{\partial V_T}{\partial m_q}$
and $m_q\frac{\partial V_{SS}}{\partial m_q}$ . They are shown in Figs. 1, 2
and 3, respectively. The spin-spin and tensor potential derivatives are
compared to those
extracted from the one pion exchange potential (OPEP),
presented in the appendix, where $g_A$ and $m_q\frac{\partial g_A}{\partial
m_q}$ were taken from Table I.

The results for $m_q\frac{\partial V_C}{\partial m_q}$ as a function of the
relative distance of the skyrmion centers, are shown in Fig. 1. The
calculations show minimum around 1 fm, with values about -20 MeV, while at
short distance they present repulsion. The difference $V_C^{(m_\pi
=138)}-V_C^{(m_\pi =0)}$ is comparable to $m_q\frac{\partial V_C}{\partial
m_q}$, as shown in the figure.

In Fig.2, the results for $m_q\frac{\partial V_T}{\partial m_q}$ are shown.
The derivative $m_q\frac{\partial V_T^{OPEP}}{\partial m_q}$ is compared
with the corresponding one for the skyrmion-skyrmion interaction. They agree
for separations of the nucleons above 3 fm. This is expected since the OPEP
tail is present in the skyrmion-skyrmion interaction. The derivative found
with OPEP is negative at short-distances, while the product ansatz results
are positive. The shapes of the curves for the difference $V_T^{(m_\pi
=138)}-V_T^{(m_\pi =0)}$ and $m_q\frac{\partial V_T}{\partial m_q}$ are
qualitatively similar, although the tail presents a noticeable difference.
We have checked that at distances above  3 fm, 
the difference in the
potentials with finite and zero pion masses is given by the results from
OPEP, since the tail is dominated by $m_\pi $ being zero or finite.

The derivative of the spin-spin component of the skyrmion-skyrmion
interaction, $m_q\frac{\partial V_{SS}}{\partial m_q}$, is shown in Fig. 3.
The derivative of the OPEP spin-spin component for distances above 3 fm
agrees with the present calculation. As expected for distances below 2 fm,
the derivatives of the skyrmion-skyrmion potential and OPEP become quite
different. The shape of the difference $V_{SS}^{(m_\pi =138)}-V_{SS}^{(m_\pi
=0)}$ is similar to that of $m_q\frac{\partial V_{SS}}{\partial m_q}$ .
However at distances above  3 fm, 
the difference in the potentials with
finite and zero pion masses is essentially dominated by the results from
OPEP, which explains the origin in the differences found.

\section{ Results for The Nuclear Sigma Term}

One way to extract the nuclear sigma term is from the low energy
pion-nucleus scattering analysis. In fact, in a recent work of Brown and Rho%
\cite{BR96} it is suggested the possibility of obtaining information about
the quark condensate from pionic atoms.

The NN potential extracted from the product ansatz is not realistic in the
sense that it does not reproduce the central attraction in the mid-range.
This cannot be ignored since it means that there is no wave function for
the bound state in this potential. Then we decided to give up consistency
and from the derivative of the calculated potential, we can estimate its
contribution to the nuclear sigma term using charge densities from the
electron elastic scattering\cite{JVV74}.

In the particular case of the deuteron, we calculated 
\begin{equation}
\Sigma _{^2\text{H}}^V=\left\langle \psi _{^2\text{H}}\right| m_q\frac{%
\partial V}{\partial m_q}\left| \psi _{^2\text{H}}\right\rangle
\end{equation}
using the normalized wave function\cite{EW88}. 
\begin{equation}
\psi _{^2\text{H}}=\psi _{J=1,M}=\left( 4\pi \right) ^{-1/2}\left\{ \frac{%
u(r)}r+\frac{w(r)}r\frac 1{\sqrt{8}}S_{12}(\hat r)\right\} \chi _{1M}.
\end{equation}
where $u$ and $w$ are the s-wave ($^3$S$_1$) and  d-wave ($^3$D$_1$) from
the deuteron wave function obtained with the Reid potential\cite{Rei68}. The
normalized wave function is such that $\int_0^\infty dr\left( u^2+w^2\right)
=1$. The spin component $\chi _{1M}$ $\equiv \left| 1M\right\rangle $ is
such that $\chi _{1M}\chi _{1M}^{\dagger }=1$ and $r=|r_1-r_2|$ is the
relative distance between the nucleons.

The Skyrme model offers naturally forces of two, three and more bodies.
Nevertheless, it is still a good approximation to neglect them all in the
nuclear sigma term except the contributions from the two-body potential.
Assuming a modified Hartree approximation  such that it has the correct
number of pairs, we have 
\begin{equation}
\Sigma _A^V=\frac{A(A-1)}2\int d^3{\bf r}_1\rho ({\bf r}_1)\int d^3{\bf r}%
_2\rho ({\bf r}_2)m_q\frac{\partial V(\left| {\bf r}_1-{\bf r}_2\right| )}{%
\partial m_q} \ , \label{signu}
\end{equation}
where $\int d^3{\bf r}\rho ({\bf r})=1$ and $A$ is the mass number.

For isoscalar nuclei heavier than the deuteron we neglected the spin-spin and
tensor components of the derivative of the potential, considering that
they are more
than one order of magnitude smaller than the central part and the
isospin is averaged in the matrix element (which yields zero in the Hartree
approximation).
Then only the derivative of the central potential
enters in the estimate of $\Sigma _A^V$.

We present in Table II the results of $\Sigma _A^V$ using (\ref{signu}) and
the densities obtained from elastic electron scattering \cite{JVV74}. In
this first approach, we assumed the same proton and neutron density
distributions. The deuteron results were included in Table II. In the case
of the deuteron the contribution from the NN interaction is $-$3.60MeV . We
expected that we could roughly count the number of NN pairs for light nuclei
in $\Sigma _A^V$. For instance, in $^4$He we can count 6 pairs, then we
should have 6$\times (-3.60)$MeV=$-$21MeV that is a lower value (in modulus)
than the calculated $-34.5$MeV. A possible explanation is that the deuteron
is ''big'' and its NN interaction is on an average weaker than in $^4$He.
Triton has 3 pairs and we estimate the contribution of the potential to the
nuclear sigma term as being between 3$\times \Sigma _{^2\text{H}}^V=10.8$MeV
and 3$\times \Sigma _{^4\text{He}}^V/6=17.2$MeV. Averaging we find a value
of 14MeV for triton. The same should work for $^3$He. For larger values 
of $A$, $\Sigma _A^V$ remains  always
smaller than the number of pairs$\times \Sigma _{^2%
\text{H}}^V$ or the number of pairs$\times \Sigma _{^4\text{He}}^V/6$, because
the NN interaction has a range around 1 fm.
Also we observe that $\Sigma _A^V/A$ is
always growing with $A$ except for the $^{197}$Au and $^{208}$Pb. This can
be explained by considering that lead have lower density distribution at
mid-range than gold, as we checked.

In Table II, we also included the kinetic energy contribution to the sigma
term. The kinetic energies from the deuteron, $^3$He and $^4$He were
extracted from the work of Schiavilla et al\cite{Sch85}. In the heavier
nuclei we took the average kinetic energy of nonrelativistic nucleons in
nuclear matter, given by $\overline{E}_K/A=\frac 3{10}(2)\varepsilon _f.$ 
At $\rho
=0.17 $fm$^{-3},$ $k_f=1.36$fm$^{-1},\varepsilon _f$ =38MeV and we arrive at 
$\overline{E}_K/A=22.8$MeV. The values of $\Sigma _A^{V+K}\equiv \Sigma
_A^V+\Sigma _A^K$ are also presented and they will be used in the
evaluation of pion-nucleus optical potential.

We noticed that the calculation of the sigma term from the NN potential
using the Gell-Mann, Oakes and Renner relation(GOR) and taking the
derivative with respect to the pion mass, produced very different results
from those calculated considering only the symmetry breaking term (according
to Adkins and Nappi \cite{Adkins}). This occurs because there is an enormous
contribution from the fourth order term. We also observed that it is in the
mid-range(1fm-2fm) that the potential NN gives the biggest contribution to
the sigma term.

\section{The nuclear sigma term in the pion-nucleus s-wave potential}

Considering that in the last section we got an estimate of the contribution of
the NN potential to the nuclear sigma term, we can also 
use it in the calculation of
the s-wave pion-nucleus potential. The interaction
Lagrangian between the pion and the nucleus which is originated by the
nuclear sigma term($\Sigma _A)$ is given by

\begin{equation}
{\cal L}_{_{\text{int}}}^\Sigma =\frac{m_q}{2f_\pi ^2}\left\langle \psi
_A\right| :\bar q(x)q(x):\left| \psi _A\right\rangle {\bf \pi }^2.
\end{equation}
We can approximate 
\begin{equation}
m_q\left\langle \psi _A\right| :\bar q(x)q(x):\left| \psi _A\right\rangle
\simeq \frac{\Sigma _A}A\rho ({\bf r)}
\end{equation}
with $\int d^3{\bf r}\rho ({\bf r)=}A.$ The nuclear sigma term is separated
in its contributions of individual nucleons, potential and kinetic energy; $%
\Sigma _A=A\Sigma _N+\Sigma _A^V+\Sigma _A^K$ (see Eq. \ref{sign}).

The component of the interaction Lagrangian
 which originates the s-wave potential $%
(U^{l=0}({\bf r}))$ can be written as 
\begin{equation}
{\cal L}_{_{\text{int}}}^{l=0}=-\frac{\Pi ^{l=0}({\bf r})}2{\bf \pi }^2
\ , \end{equation}
where the  s-wave self energy is given by 
\begin{equation}
\Pi ^{l=0}({\bf r})=-4\pi (1+m_\pi /M_N)[b_{0_{\text{eff}}}^{free}+%
\frac{\Sigma _A^{V+K}}{4\pi f_\pi ^2A}(1+\frac{m_\pi }{M_N})^{-1}]\rho ({\bf %
r).}  \label{7.35}
\end{equation}

We have separated $\Sigma _A^{V+K}$ and $b_{0_{\text{eff}}}^{free}$
contributions in the self energy $\Pi ^{l=0}({\bf r}),$ Eq. (\ref{7.35}).
This last one should contain all the mechanisms of the pion-nucleus interaction
not included in $\Sigma _A^{V+K}.$ The normalization was chosen such that $%
b_{0_{\text{eff}}}^{free}$ has the same normalization as the $\pi N$
scattering length.

Introducing the self energy, $\Pi ^{l=0}({\bf r}),$
in Klein-Gordon equation for the pion \cite{EW88,EK80}, 
using $\omega ^2=k^2+m_\pi ^2$ and dividing by  $2m_\pi $, we get 
\begin{equation}
\left[ -\frac{{\bf \nabla }^2}{2m_\pi }+\frac{2\omega V_{Coul}}{2m_\pi }-%
\frac{V_{Coul}^2}{2m_\pi }+\frac{\Pi ^{l=0}({\bf r})}{2m_\pi }\right]
\varphi _a({\bf r})=\frac{k^2}{2m_\pi }\varphi _a({\bf r}).  \label{7.36}
\end{equation}
In this way we have obtained the pion Schr\"odinger equation, where we
identify the potential 
\begin{equation}
U^{l=0}({\bf r})=\frac 1{2m_\pi }\Pi ^{l=0}({\bf r}).  \label{7.37}
\end{equation}

The part of the s-wave optical potential from the term $\Sigma _A^{V+K}$
is given by 
\begin{equation}
U_{\Sigma _A^{V+K}}^{l=0}({\bf r})=-\frac{\Sigma _A^{V+K}}{2m_\pi f_\pi ^2A}%
\rho ({\bf r),}
\end{equation}
where Eq.s (\ref{7.35}) and (\ref{7.37}) were used. Then, if $\Sigma
_A^{V+K} $ is positive, it will produce an attractive contribution to the
optical potential and repulsive otherwise.

The scattering amplitude in the long-wave length limit and in the Born
approximation is given by

\begin{equation}
f({\bf k})=-\frac{\mu _\pi }{2\pi }\int d^3{\bf r}e^{i{\bf k.r}}U^{l=0}({\bf %
r}).  \label{7.39}
\end{equation}
The nucleus recoil is considered in Eq.(\ref{7.39}) by the reduced mass
factor $\mu _\pi =m_\pi M_A/(m_\pi +M_A)$, that 
corresponds to the addition of the nonrelativistic 
kinetic energy of the target in the pion Schr\"odinger equation
(\ref{7.36}). The pion-nucleus scattering length in Born approximation is
given by $a=f(0).$

Although we have discussed the Born approximation in general, it only will
be used for the deuteron, $^3$He and $^4$He, when discussing experimental
data. For heavier nuclei we will compare the calculations to the Seki-Masutani
parameter $b_0^{\text{SM}}$\cite{Sek88}
 extracted from experimental data analysis and
defined as the coefficient of the $\rho {\bf (r)}$ term in the optical
potential, 
\begin{equation}
2m_\pi U^{l=0}({\bf r})=-4\pi (1+m_\pi /M_N)b_0^{\text{SM}}\rho ({\bf r}).
\label{7.41}
\end{equation}
Comparing the equations (\ref{7.35}),(\ref{7.37}) and (\ref{7.41}), we get 
\begin{equation}
b_0^{\text{SM}}=b_{0_{\text{eff}}}^{free}+b_0^{V+K},  \label{b0sm}
\end{equation}
where, 
\begin{equation}
b_0^{V+K}=\frac{\Sigma _A^{V+K}}{4\pi f_\pi ^2A}\left( 1+\frac{m_\pi }{M_N}%
\right) ^{-1}
\end{equation}
which is the contribution of the nuclear sigma term from the NN potential
plus the kinetic energy to $b_0^{\text{SM}}.$

The isoscalar scattering length $b_{0_{\text{eff}}}^{free}$ was obtained
using the experimental value of the $\pi d$ scattering length ($%
\mathop{\rm Re}
a_{\pi d}=-0.0264(11)m_\pi ^{-1}$) \cite{Cha95}, 
\begin{equation}
b_{0_{\text{eff}}}^{free}=\left( 1+\frac{m_\pi }{M_N}\right) ^{-1}\left[ (1+%
\frac{m_\pi }{M_d})\frac{%
\mathop{\rm Re}
a_{\pi d}}2-\frac{\Sigma _d^{V+K}}{8\pi f_\pi ^2}\right] ,  \label{7.44}
\end{equation}
with $m_\pi =138$MeV, $M_N=938$MeV, $M_d$ is the deuteron mass, $f_\pi =93$%
MeV. We used $\Sigma _{^2\text{H}}^{V+K}=-$4.46MeV, extracted from Table II,
for $f_\pi =93$ MeV and $e=$4.0 . These parameters adjust the $\pi N$ 
p-wave scattering \cite{Holz}. Substituting these values in (\ref{7.44}) we
get 
\begin{equation}
b_{0_{\text{eff}}}^{free}=-\text{0.0099 (}m_\pi ^{-1}).
\end{equation}

This value is consistent with $-0.0077(11) \ m_\pi ^{-1}$ got from Sigg et al 
\cite{Sig95} in the isospin symmetry hypothesis and also with the value
$-0.0083(38) \ m_\pi ^{-1}$ from Koch\cite{Koc86}.

It has been noted that the scattering length for bound nuclei is shifted
from the scattering length predicted by the theory for free nucleons%
\footnote{%
>From Kluge\cite{Klu91}: ''The conclusion is that this repulsion cannot be
obtained by an iteration of $\pi N$ interaction, but rather represents an
independent feature of the interaction of pions with bulk nuclear matter. An
attempt to explain this as a binding effect of the nucleons has been
undertaken in the framework of relativistic mean field theory of the nucleus
(Birbair et al 1983 J.Phys.G: Nucl.Phys 9 1473; Goudsmit et al 1989 Preprint
ETHZ-IMP RP-90-03 ETH Z\"{u}rich)''}. This suggests that the cause for this
shift is precisely the nuclear sigma term and kinetic energy contribution.
The value of the isoscalar $\pi N$ scattering length parameter $b_0$ is
controversial and is not our purpose to calculate it here. Then, we obtained
the effective isoscalar $b_{0_{\text{eff}}}^{free}$ originated from
experimental pion-deuteron scattering length that contains the single and
double scattering processes not included in $\Sigma _{^2\text{H}}^{V+K}.$

We constructed Table III
using the results of the potential and kinetic energy 
contributions to the sigma term in  Eq.(\ref{b0sm}) . The
experimental results were extracted from pionic atoms and pion-nucleus
scattering analysis \cite{Sek83,Sek88} .

In the particular case of $^3$He, $b_{0_{\text{eff }}}^{\text{(exp)}}$ was
evaluated subtracting the isovector contribution, i.e., considering only
single scattering,

\begin{equation}
b_{0_{\text{eff }}\pi ^{-}A}^{\text{(exp)}}=\frac 1A\left[ \left( 1+\frac{%
m_\pi }{M_N}\right) ^{-1}(1+\frac{m_\pi }{AM_N})%
\mathop{\rm Re}
a_{\pi ^{-}A}-(N-Z)b_1\right]  \label{b0b12}
\end{equation}
and using the experimental values $%
\mathop{\rm Re}
a_{\pi ^{-}\text{ }^3\text{He}}=$0.056$(6)$ \cite{EW88} and $b_1$=$-$%
0.0962(7) extracted from \cite{Sig95} we got $b_{0_{\text{eff }}}^{\text{%
(exp)}}(^3$He)$\approx -$0.015. The experimental result for $^{12}$C was
obtained by the Seki-Masutani analysis and also for $^{16}$O extrapolated
 from $\pi ^{-}$ scattering data at 29MeV and 50 MeV \cite{Sek83}. The
experimental result for $^{40}$Ca was found in the Masutani-Seki analysis%
\cite{Sek88} extrapolated from $\pi ^{-}$ scattering data at 25MeV to zero
energy. In the cited work \cite{Sek83}, it is also found that the 
pion interacts with the nucleus
at an effective density $\rho _e=\rho _N/2$. The empirical value of $-$0.0304%
$m_\pi ^{-1}$ was obtained by Salcedo et al\cite{Sal95} averaging values
obtained from the analysis of pionic atom and pion-nucleus scattering data.

Our evaluation of the nuclear sigma term from the NN potential generates a
repulsion in the pion-$^{12}$C optical potential in the central region which
is approximately $U_\Sigma ^{V+K}({\bf r})=-\frac{\Sigma _A^{V+K}}{2m_\pi
f_\pi ^2A}\rho ({\bf r})
\approx 8$ MeV that is roughly of the same order  necessary to fit
the pionic atoms ( $\approx $15 to 30 MeV$)$ and is ''weakly dependent of $A"$
\cite{Klu91}.

\section{Conclusions}

We have calculated the derivative of the skyrmion-skyrmion interaction with
respect to the quark mass, $m_q\frac{\partial V_{NN}}{\partial m_q}$, in the
quantized product ansatz. We have related it to $m_\pi ^2\frac{\partial
V_{NN}}{\partial m_\pi ^2}$ using the Gell-Mann, Oakes and Renner relation,
which made the calculation possible. We compared it to the OPEP calculation,
and they agreed for distances above 3 fm. The matrix element of above
derivative of $V_{NN}$ in the nuclear state is part of the nuclear sigma
term. The magnitudes of the quark mass contribution to the nuclear potential
were found appreciable. We calculated the contribution to the nuclear
sigma term arising from the nuclear interaction, 
assuming that the nuclear density in
finite nuclei is equal to the experimental charge density conveniently
normalized.

The contribution from the NN potential to the sigma term was incorporated in
the effective isoscalar scattering length of the pion-nucleus optical
potential. There was a remarkable agreement with the empirical values
extracted from the analysis of pionic atoms and scattering experimental data.
We showed that it is possible to explain the repulsion in the isoscalar
channel of the pion-nucleus optical potential, if we take into account 
in the nuclear sigma term the contribution of the NN potential. 

In nuclear matter, we should include the contribution of the NN potential 
($\Sigma_A^V$) in the nuclear sigma term   
in the form $\Sigma _A=A\Sigma _N+\Sigma
_A^V+\Sigma _A^K.$ This correction alters the value of the condensate $%
\left\langle \bar qq\right\rangle _\rho /\left\langle \bar qq\right\rangle
_{vac}=1-\rho \Sigma _A/Am_\pi ^2f_\pi ^2,$ and as $\Sigma _A^V$ is negative
there is a tendency to a drastic reduction of $\Sigma _A$. Then the
condensate at normal nuclear density is about $\left\langle \bar q%
q\right\rangle _{\rho _N}/\left\langle \bar qq\right\rangle _{vac}\approx $
0.95, and by this way it does not decrease as much as it was calculated
before. More definitive conclusions about the quark mass dependence of the
NN potential, should be pursued in exact evaluations of the
skyrmion-skyrmion interaction, that is beyond the present work.

{\bf Acknowledgments}

AG would like to thank M.R. Robilotta for the introduction
 to the Skyrme model. We also thank M.R. Robilotta and J.M.Eisenberg for 
valuable discussions on this subject. We  express our gratitute to 
Vera B. Canteiro for reading the manuscript. This work was supported by
FAPESP and CNPq of Brazil.

{\bf Appendix}

The nonrelativistic reduction of the 
one pion exchange interaction is given by\cite{Ro97} 
\begin{eqnarray}
V^{OPEP} &=&\frac{g_A^2}{4\pi }\left( \frac{m_\pi }{2f_\pi }\right) ^2(\vec{%
\tau}_1.\vec{\tau}_2)\left[ \frac 13\left( {\vec{\sigma}}_1.{\vec{\sigma}}%
_2\right) +\left( \frac 1{\left( m_\pi r\right) ^2}+\frac 1{m_\pi r}+\frac 13%
\right) S_{12}\right] \frac{e^{-m_\pi r}}r  \nonumber \\
&=&(\vec{\tau}_1.\vec{\tau}_2)\left[ V_{SS}^{OPEP}\left( {\vec{\sigma}}_1.{%
\vec{\sigma}}_2\right) +V_T^{OPEP}S_{12}\right] \text{ ,}  \eqnum{a.1}
\end{eqnarray}

The derivatives of the potentials with respect to the square of the pion
mass can be simply stated as 
\begin{equation}
m_\pi ^2\frac{\partial V_{SS}^{OPEP}}{\partial m_\pi ^2}=\frac 1{12\pi }%
\left( \frac{m_\pi }{2f_\pi}\right) ^2\left\{ 2g_{A}\frac{\partial g_{A} }{%
\partial m_\pi ^2}m_\pi ^2+g_{A}^2\left( 1-\frac{m_\pi r}2\right)\right\} 
\frac{e^{-m_\pi r}}r  \eqnum{a.2}
\end{equation}
and 
\begin{eqnarray}
m_\pi ^2\frac{\partial V_T^{OPEP}}{\partial m_\pi ^2} &=&\frac 1{4\pi }%
\left( \frac{m_\pi }{2f_\pi}\right) ^2\left\{ 2g_{A}\frac{\partial g_{A}} {%
\partial m_\pi ^2}m_\pi ^2\left( \frac 1{\left( m_\pi r\right) ^2}+\frac 1{%
m_\pi r}+\frac 13\right) -\frac{g_A^2}6\left( 1+m_\pi r\right) \right\} 
\frac{e^{-m_\pi r}}r.  \eqnum{a.3}
\end{eqnarray}

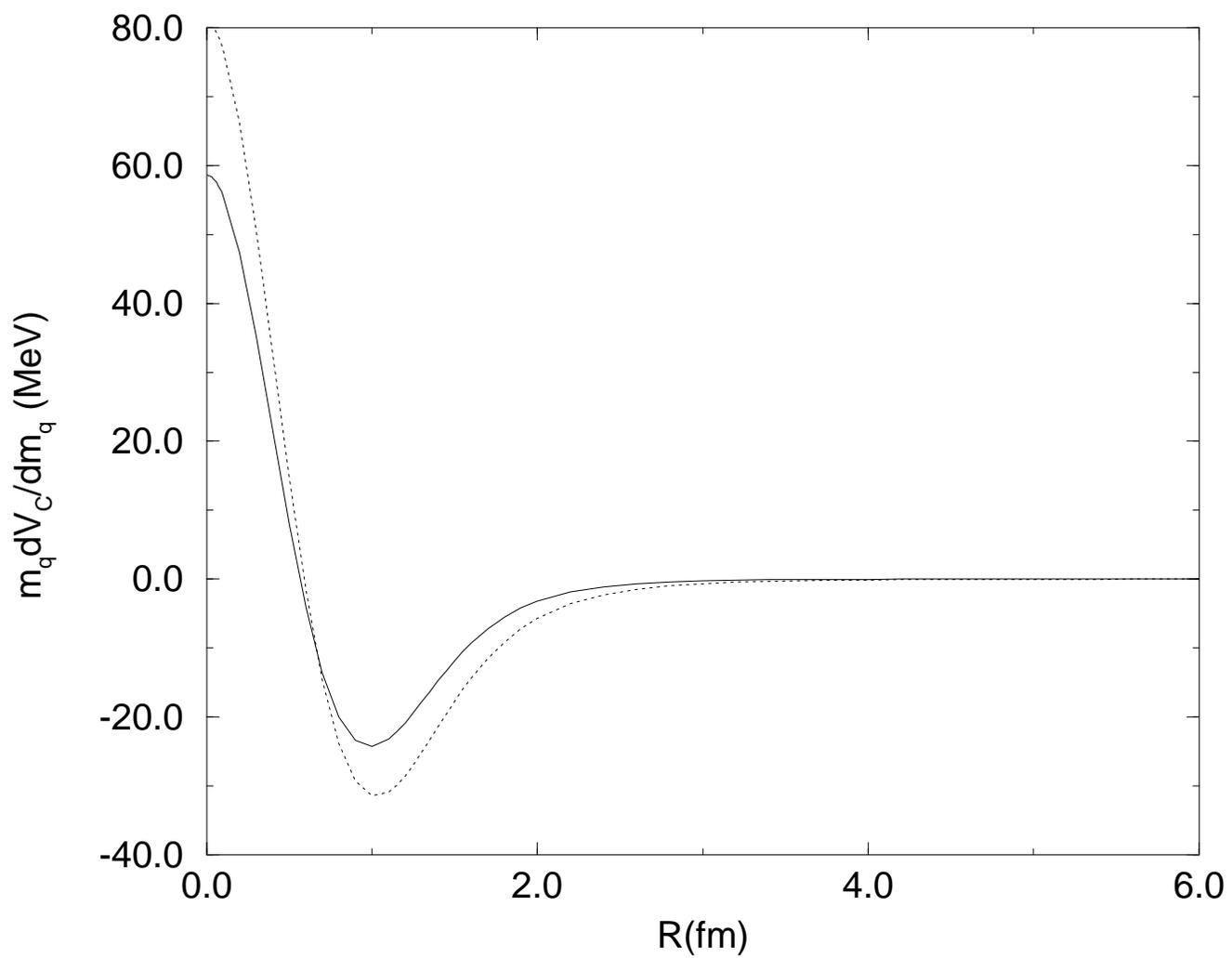
\begin{figure}[tbp]
\caption{Derivative of the central component of the NN interaction, $m_q%
\frac{\partial V_C}{\partial m_q}$ as a function of the nucleon-nucleon
relative distance in the Skyrme model. Solid curve corresponds to $m_q\frac{%
\partial V_C}{\partial m_q}$ with $f_\pi =93$MeV and $e=4.0$. Dotted curve
represents $V_C^{(m_\pi=138)}-V_C^{(m_\pi=0)}$. }
\label{fig1}
\end{figure}

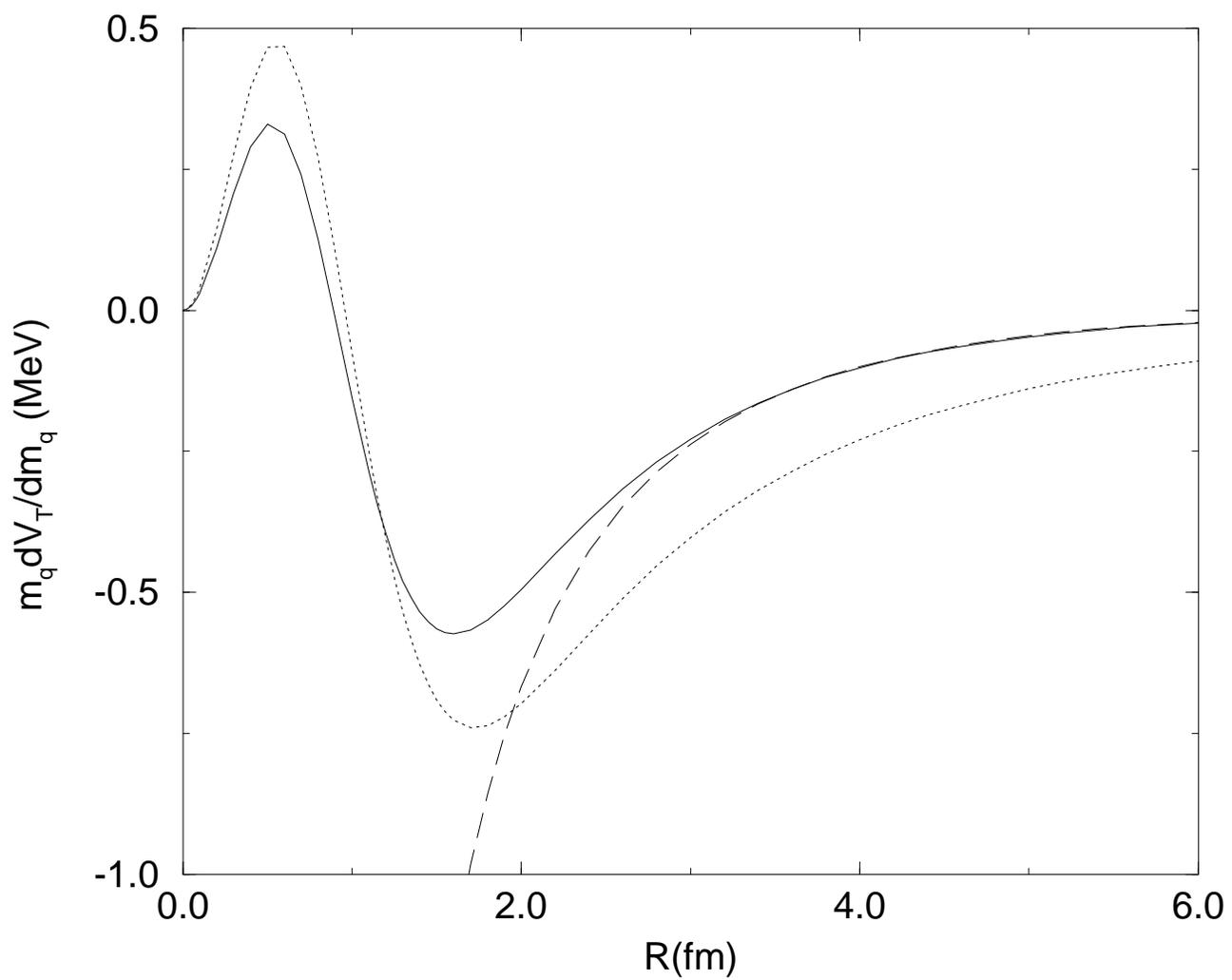
\begin{figure}[tbp]
\caption{Derivative of the tensor component of the NN interaction, $m_q\frac{%
\partial V_T}{\partial m_q}$ as a function of the nucleon-nucleon relative
distance in the Skyrme model. Solid curve corresponds to $m_q\frac{\partial
V_T}{\partial m_q}$ with $f_\pi =93$MeV and $e=4.0$. The dashed curve
corresponds to the OPEP results. Dotted curve represents $V_T^{(m_%
\pi=138)}-V_T^{(m_\pi=0)}$.}
\label{fig2}
\end{figure}

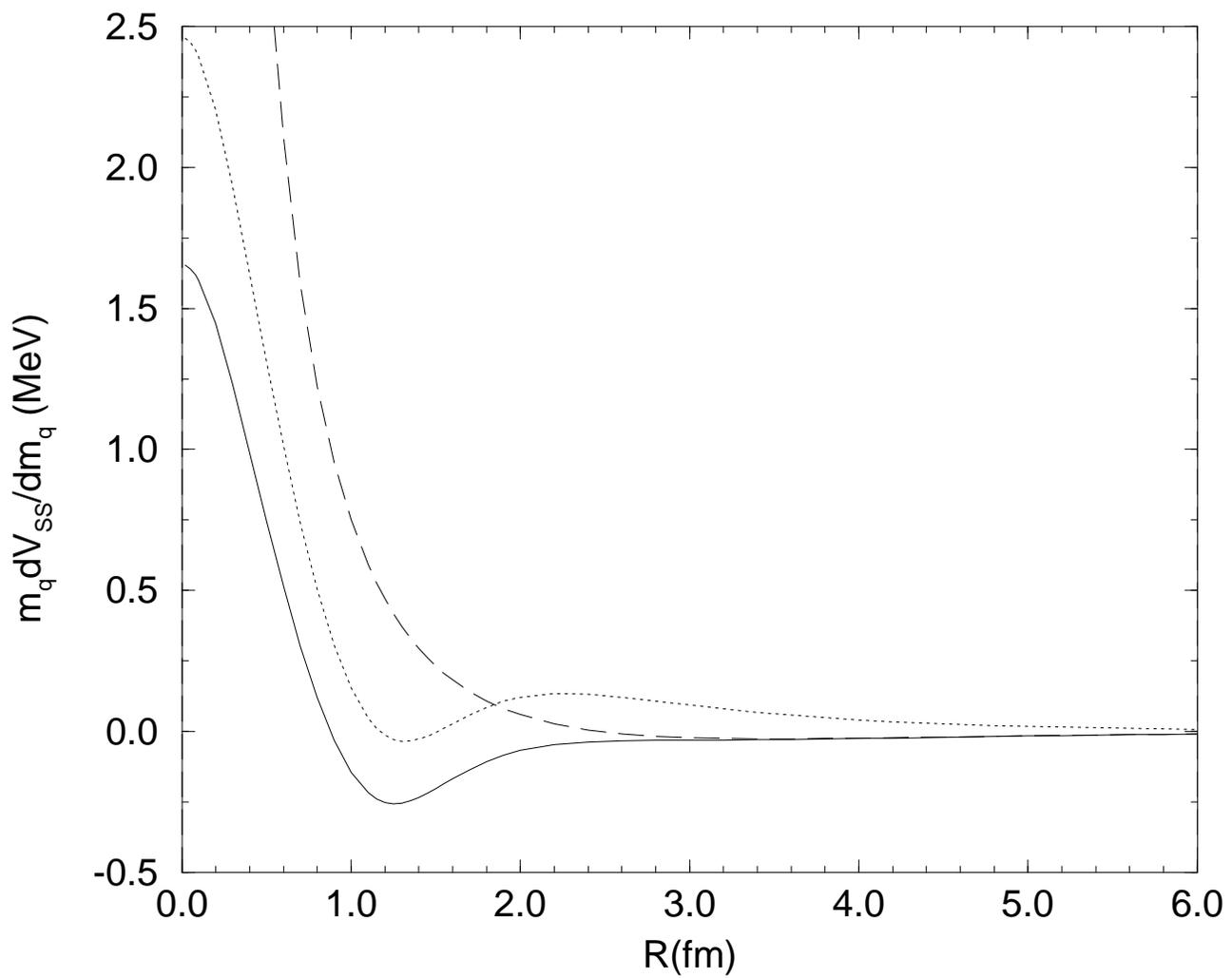
\begin{figure}[tbp]
\caption{Derivative of the spin-spin component of the NN interaction, $m_q%
\frac{\partial V_{SS}}{\partial m_q}$ as a function of the nucleon-nucleon
relative distance in the Skyrme model. The labels of the curves are the same
as in Fig.2. }
\label{fig3}
\end{figure}

\newpage
\mediumtext

\begin{table}[tbp]
\caption{Nucleon observables ($O$) and the corresponding derivatives with
respect to the quark mass.}
\begin{tabular}{|c|c|c||c|c|}
& \multicolumn{2}{c||}{$f_\pi =54$ MeV $e=4.84$} & \multicolumn{2}{c|}{$%
f_\pi =93$ MeV $e=4.0$} \\ \hline
& $O$ & $m_q\frac{\partial O}{\partial m_q }$ & $O$ & $m_q\frac{\partial O}{%
\partial m_q }$ \\ \hline
$M_N$ & 938 MeV & 52.2 MeV & 1818MeV & 59.6 MeV \\ \hline
$g_A$ & 0.65 & $-0.064$ & 1.01 & $-0.069$ \\ \hline
$g_{\pi NN}$ & 11.88 & $-0.022$ & 20.35 & $-0.33$ 
\end{tabular}
\end{table}

\begin{table}[tbp]
\caption{Contributions of the derivative of the NN potential and average
kinetic energy to the nuclear sigma term. Calculations performed for the
Skyrme Model with $f_\pi $=93MeV and e=4.0. The upper label in the nucleus
symbol denotes the parametrization of the densities according to de Jager,
de Vries and de Vries[21]. }
\begin{tabular}{|c|c|c|c|c|c|}
& $\bar{E}_K$(MeV) & $\Sigma _A^K$(MeV) & $\Sigma _A^V$(MeV) & $%
\Sigma_A^{V+K}$ (MeV) & $\Sigma_A^{V+K}/A$ (MeV) \\ \hline
$^2$H & 17.8 & -0.86 & -3.60 & -4.46 & -2.23 \\ \hline
$^3$H & 46.1 & -2.3 & $\approx $-14 & $\approx $-16 & $\approx $-5.4 \\ 
\hline
$^3$He & 46.1 & -2.3 & $\approx $-14 & $\approx $-16 & $\approx $-5.4 \\ 
\hline
$^4$He$^{\text{(3pF)}}$ & 106 & -5.09 & -34.5 & -39.6 & -9.9 \\ \hline
$^{10}$B$^{\text{(h.o.)}}$ & 228 & -10.9 & -107 & -118 & -11.8 \\ \hline
$^{12}$C$^{\text{(3pF)}}$ & 274 & -13.1 & -157 & -170 & -14.2 \\ \hline
$^{14}$N$^{\text{(h.o.)}}$ & 319 & -15.3 & -188 & -203 & -14.5 \\ \hline
$^{16}$O$^{\text{(3pF)}}$ & 365 & -17.5 & -220 & -238 & -14.9 \\ \hline
$^{20}$Ne$^{\text{(2pF)}}$ & 456 & -21.8 & -268 & -290 & -14.5 \\ \hline
$^{24}$Mg$^{\text{(3pF)}}$ & 547 & -26.2 & -388 & -414 & -17.2 \\ \hline
$^{32}$S$^{\text{(3pF)}}$ & 730 & -35.0 & -557 & -592 & -18.5 \\ \hline
$^{40}$Ca$^{\text{(3pG)}}$ & 912 & -43.8 & -727 & -771 & -19.3 \\ \hline
$^{56}$Fe$^{\text{(3pG)}}$ & 1273 & -61.1 & -1144 & -1205 & -21.5 \\ \hline
$^{118}$Sn$^{\text{(3pG)}}$ & 2690 & -129 & -2763 & -2892 & -24.5 \\ \hline
$^{148}$Sm$^{\text{(2pF)}}$ & 3374 & -162 & -3556 & -3718 & -25.1 \\ \hline
$^{197}$Au$^{\text{(2pF)}}$ & 4492 & -216 & -5187 & -5403 & -27.4 \\ \hline
$^{208}$Pb$^{\text{(3pG)}}$ & 4742 & -227 & -5274 & -5501 & -26.4 \\ \hline
$\rho ($0.17fm$^{-3})$ & 22.8$A$ & -1.09$A$ & -38.0$A$ & -39.1$A$ & -39.1
\end{tabular}
\end{table}

\begin{table}[tbp]
\caption{Results for the isoscalar parameter of the optical potential
including the potential sigma term contribution in Eq.(\ref{b0sm}). The $%
^{12} $C and $^{16}$O experimental values were estimated from the analysis
described in Ref.[10] and for $^{40}$Ca from the analysis of Ref.[11]. The
value of $b_{0_{\text{eff}}}^{free} $ 
was obtained from the experimental value of
the $\pi$-d scattering length [26] 
($b_{0_{\text eff}}^{free} = -0.0099 \ (m_\pi^{-1}$)) .}
\begin{tabular}{|c|c|c|c|}
& $b_0^{V+K}$ & $b_0^{V+K}+b_{0_{\text{eff}}}^{free}$ & $b_{0_{\text{eff}}}^{{%
({\text{exp}})}}$ \\ \hline
$^2$H & -0.0025 & -0.0123 & -0.0123(3) \\ \hline
$^3$H & -0.0063 & -0.0162 & - \\ \hline
$^3$He & -0.0063 & -0.0162 & $\approx $-0.015 \\ \hline
$^4$He$^{\text{(3pF)}}$ & -0.0110 & -0.0209 & -0.0212(5) \\ \hline
$^{10}$B$^{\text{(h.o)}}$ & -0.0131 & -0.0230 & - \\ \hline
$^{12}$C$^{\text{(3pF)}}$ & -0.0157 & -0.0256 & -0.0260 \\ \hline
$^{14}$N$^{\text{(h.o.)}}$ & -0.0160 & -0.0259 & - \\ \hline
$^{16}$O$^{\text{(3pF)}}$ & -0.0165 & -0.0264 & -0.026$^{^{(50\text{MeV)}}}$%
;-0.027$^{^{(29\text{MeV)}}}$ \\ \hline
$^{20}$Ne$^{\text{(2pF)}}$ & -0.0160 & -0.0259 & - \\ \hline
$^{24}$Mg$^{\text{(3pF)}}$ & -0.0190 & -0.0289 & - \\ \hline
$^{32}$S$^{\text{(3pF)}}$ & -0.0204 & -0.0303 & - \\ \hline
$^{40}$Ca$^{\text{(3pF)}}$ & -0.0213 & -0.0312 & -0.033$^{^{(25\text{MeV)}}}$
\\ \hline
$\rho _e=0.17$fm$^{-3}/2$ & -0.0216 & -0.0315 & -0.0304
\end{tabular}
\end{table}

\end{document}